\documentclass{article}
\usepackage{amsmath}
\usepackage{amssymb}
\newtheorem{lemma}{Lemma}
\begin{document}

\begin{titlepage}
\begin{center}
{\Large
Quantum Systems with Linear Constraints and Quadratic Hamiltonians
}
\\[1cm]
{{\large  O.Yu.Shvedov} \\[0.5cm]
{\it
Sub-Dept. of Quantum Statistics and Field Theory},\\
{\it Dept. of Physics, Moscow State University},\\
{\it 119992, Moscow, Vorobievy Gory, Russia}
}
\\[1cm]
{\bfseries \itshape
Talk given at the International Conference \\
"Symmetry in Nonlinear
Mathematical Physics", \\
Kyiv, June 20-26, 2005}
\\[0.5cm]~
\end{center}

\setcounter{page}{0}

\begin{flushright}
math-ph/0512089
\end{flushright}

\section*{Abstract}

Quantum systems  with  constraints  are  often  considered  in  modern
theoretical physcics.  All realistic field models based on the idea of
gauge  symmetry are of this type.  A partial case of constraints being
linear in coordinate and momenta operators is very important.  Namely,
when  one  applies  semiclassical  methods  to an arbitrary constraint
system,  the constraints in "general position case" become linear.  In
this paper,  different mathematicals constructions for the Hilbert space
space for the constraint system are discussed.  Properties of Gaussian
and  quasi-Gaussian wave functions for these systems are investigated.
An  analog  of  the  notion  of  Maslov  complex  germ  is  suggested.
Properties  of  Hamiltonians  being  quadratic  with  respect  to  the
coordinate and momenta operators are discussed.
The Maslov theorem (it says
that there exists a Gaussian eigenfunction of the quantum  Hamiltonian
iff  the classical Hamiltonian system is stable) is generalized to the
constrained systems.  The case of infinite number degrees  of  freedom
(constrained Fock space) is also discussed.

{\it Keywords:}
constrained systems; stability; Maslov complex germ

\footnotetext{e-mail:  olegshv@mail.ru}

\footnotetext{This work was supported by the Russian  Foundation  for
Basic Research, project 05-01-00824.
}
\end{titlepage}

\section{Introduction}

Quantum constrained  systems play an important role in modern physics.
All physical quantum field models  (electrodynamics,  standard  model,
chromodynamics) for  elementary  particle  physics are gauge theories,
examples of constrained systems \cite{Shvedov:SF}.

When one applies the Maslov complex-WKB theory 
\cite{Shvedov:Maslov1,Shvedov:Maslov2}  to
the constrained  systems  \cite{Shvedov:Shv1},  one obtains an evolution equation
with quadratic Hamiltonian  operator  and  linear  constraints.  Thus,
investigation of  quantum  constrained systems with linear constraints
and quadratic Hamiltonians is important.

Different constructions  for  Hilbert  space  for  such  systems   are
considered in  this  paper.  Properties of Gaussian and quasi-Gaussian
wave functions usually appearing  in  the  Maslov  cpmplex-WKB  theory
\cite{Shvedov:Maslov2} are  studied.  Evolution  transformation for the quadratic
Hamiltonian is investigated.

There is  a  well-known  Maslov  theorem  \cite{Shvedov:Maslov2}  that  a  linear
Hamiltonian system is stable iff the corresponding quantum Hamiltonian
operator possesses  a  Gaussian   eigenfunction.   This   theorem   is
generalized to the case of a constraint system.

In quantum  field theory,  infinite-dimensional systems usually arise.
Therefore, specific features of infinite-dimensional case will be also
studied.

\section{Constrained systems and their quantization}

For the  classical case,  specific features of constrained systems are
presented in table  1.  We  see  that  constrained  systems  are  more
complicated. Phase  space is curved (not flat),  a new notion of gauge
equivalence arise.

Different approaches  to  quantize  constrained  systems   have   been
developed. Dirac    procedure    \cite{Shvedov:Dirac}   and   refined   algebraic
quantization \cite{Shvedov:Astekar} are presented in table  2  for  the  case  of
Abelian constrained   systems.   There  is  also  a  more  complicated
approach, BRST-BFV quantization \cite{Shvedov:BRST,Shvedov:BFV}. However, for the case of
quadratic Hamiltonians  and linear constraints,  it is possible to use
simpler approaches.

\begin{table}
\caption{Classical systems without constraints vs constrained systems}
\begin{tabular}{|p{1.5cm}|p{8cm}|p{5cm}|}
\hline
& Systems without constraints
& Constrained systems
\\ \hline
Phase space
&
${\cal M} = \mathbb{R}^{2n} =
\{ (P,Q)| P,Q \in \mathbb{R}^n\}$,
$P$ --- momenta, $Q$ --- coordinates.
&
surface $\cal X$ in $\mathbb{R}^{2n}$: set of all $(P,Q)$
such that $\Lambda_a(P,Q)=0$, $a=\overline{1,k}$,
$\Lambda_a$ being constraints
\\
\hline
Evolution transformation
& $u_t^H:{\cal M} \to {\cal M}$ takes the initial  condition  for  the
Hamiltonian system
\begin{displaymath}
\dot{Q}_i = \frac{\partial H}{\partial P_i},
\quad
\dot{P}_i = - \frac{\partial H}{\partial Q_i}
\end{displaymath}
to the solution of the Cauchy problem.
For any observable $A=A(P(t),Q(t))$, one has:
\begin{displaymath}
\frac{dA}{dt} = \{A,H\} =
\sum_i \left(
\frac{\partial A}{\partial Q_i} \frac{\partial H}{\partial P_i}
-
\frac{\partial A}{\partial P_i} \frac{\partial H}{\partial Q_i}
\right).
\end{displaymath}
&
Additional requirements:  constrained  conditions   $\Lambda_a(P,Q)=0$
should conserve, so that $\{\Lambda_a,H\} = 0$  on $\cal X$.
\\ \hline
Gauge equivalence
& No gauge equivalence
& For the first-class constraints \cite{Shvedov:Dirac},
$\{\Lambda_a,\Lambda_b\} = 0$ on $\cal X$,  $\Lambda_a$ are generators
of gauge     transformations.     Classical     states     $X$     and
$u_{\tau}^{\Lambda_a} X$ are set to be equivalent.
\\
\hline
\end{tabular}
\end{table}

\begin{table}
\caption{
Dirac quantization   \cite{Shvedov:Dirac}   and  refined  algebraic  quantization
(algebraic approach) \cite{Shvedov:Astekar} for Abelian constrained systems.
}
\begin{tabular}{|p{2cm}|p{5cm}|p{7cm}|}
\hline
& Systems without constraints
& Constrained systems
\\ \hline
States and observables
& Wave functions $\psi(\xi)$, $\xi \in \mathbb{R}^n$.
Observables: $p,q  \mapsto  \hat{p}_i  =  - i \frac{\partial}{\partial
\xi_i}, \hat{q}_i = \xi_i$.
& Wave functions $\psi_D(\xi)$ (Dirac approach)
or $\psi_A(\xi)$ (algebraic approach).
For the case of Dirac aproach,
additional condition $\hat{\Lambda}_a \psi_A = 0$ is imposed.
Correspondence of $\psi_A$ and $\psi_D$:
$\psi_D = \prod_a 2\pi \delta(\hat{\Lambda}_a) \psi_A$.
\\ \hline
Inner product
&
\begin{displaymath}
(\psi,\psi) = \int d\xi \psi^*(\xi) \psi(\xi)
\end{displaymath}
&
\begin{displaymath}
\left<\psi_A,\psi_A\right> = (\psi_A,
\prod_a 2\pi \delta(\hat{\Lambda}_a) \psi_A).
\end{displaymath}
\\ \hline
Evolution
&
$\psi \mapsto e^{-i\hat{H}t} \psi$.
&
Additional requirement:  Dirac  condition  and  inner  prosuct  should
conserve; this means $[\hat{\Lambda}_a,\hat{H}]=0$.
\\
\hline
Gauge equivalence
& No gauge equivalence
& States
\begin{displaymath}
\psi_A \mapsto e^{-i\lambda^a \hat{\Lambda}_a}\psi_A
\end{displaymath}
are set to be gauge equivalent.
\\
\hline
\end{tabular}
\end{table}

\section{Constructions of Hilbert space}

\subsection{Simple example with one unphysical degree of freedom}

{\bf 1}.
Consider the  simplest  example  of  a  constrained  system  with  one
unplysical degree of  freedom  (coordinate  $\hat{q}_1$  and  momentum
$\hat{p}_1$) and one constraint
\begin{equation}
\hat{\Lambda}_1 = \hat{p}_1.
\label{Shvedov:e1a}
\end{equation}
For both Dirac and algebraic approaches to  quantize  this  constraint
system, states  are  viewed as wave functions of one argument $\xi_1$,
$\psi_D(\xi_1)$ and $\psi_A(\xi_1)$.

For the algebraic  approach,  wave  function  $\psi_A(\xi_1)$  may  be
arbitrary. However, the inner product is modified,
\begin{gather}
\left<\psi_A,\psi_A\right> = 2\pi (\psi_A,\delta(\hat{p}_1)\psi_A) =
\int d\alpha (\psi_A, e^{i\alpha \hat{p}_1} \psi_A) 
\nonumber \\
=
\int d\xi_1 d\alpha \psi_A^*(\xi) \psi_A(\xi+\alpha) =
\left|\int d\xi_1 \psi_A(\xi_1) \right|^2.
\label{Shvedov:e1}
\end{gather}
Thus, one  can  say  that  states  $\psi_A^{I}$  and $\psi_A^{II}$ are
equivalent (this is a quantum gauge equivalence!) iff
\begin{displaymath}
\psi_A^I \sim \psi_A^{II} \Leftrightarrow
\int d\xi_1 \psi_A^{I}(\xi_1) =
\int d\xi_1 \psi_A^{II}(\xi_1).
\end{displaymath}
Therefore, the  equivalence  classes  of  wave  functions   $[\psi_A]$
("quantum gauge  orbits")  can  be  identified  with  complex  numbers
$\overline{\psi} =  \int  d\xi_1  \psi_A(\xi_1)$.  As  it   has   been
expected, the physcial Hilbert space of equivalence classes is trivial
(one-dimensional).

For the  Dirac  approach,  states  are  viewed   as   wave   functions
$\psi_D(\xi_1)$ satisfying additional constraint condition
\begin{displaymath}
\hat{p}_1 \psi_D = - i \frac{\partial \psi_D}{\partial \xi_1} = 0.
\end{displaymath}
It means  that  $\psi_D  =  const$.  There  is a well-known problem of
introducing an inner product in the Dirac approach.  A naive procedure
is to set $(\psi_D,\psi_D) = \int d\xi_1 |\psi_D(\xi_1)|^2$.  It fails
since the integral diverges. Therefore, one should be more careful.

To find an expression for  the  inner  product,  one  should  use  the
formula for correspondence of  Dirac and algebraic approaches,
\begin{displaymath}
\psi_D(\xi) = 2\pi \delta(\hat{p}_1) \psi_A(\xi_1) =
\int d\alpha \psi_A(\xi_1+\alpha)  =  \int  d\alpha  \psi_A(\alpha)  =
const.
\end{displaymath}
Making use of formula ~\eqref{Shvedov:e1}, one finds that
\begin{equation}
\left< \psi_D,\psi_D \right> = |\psi_D(\xi_1)|^2
\label{Shvedov:e2}
\end{equation}
for arbitrary $\xi_1$. Expression ~\eqref{Shvedov:e2} does not depend on particular
choice of $\xi_1$, since $\psi_D$ should be $\xi_1$-independent.

{\bf 2.}   Another   simple   example   of  a  constrained  system  is
one-dimensional system with the constraint of the form
\begin{equation}
\hat{\Lambda}_1 = \hat{q}_1.
\label{Shvedov:e1b}
\end{equation}
Analogously, one  finds  that  $\psi_A(\xi_1)$  may  be  an  arbitrary
function, but the inner product is to be modified,
\begin{displaymath}
\left<\psi_A,\psi_A\right> = 2\pi |\psi_A(0)|^2,
\end{displaymath}
so that equivalence classes $[\psi_A]$  are  identified  with  numbers
$\overline{\psi} = \sqrt{2\pi} \psi_A(0)$.

In the  Dirac  approach,  wave  functions  $\psi_D(\xi_1)$  obeys  the
condition $\hat{q}_1 \psi_D=0$, or
\begin{displaymath}
\psi_D(\xi_1) =  const  \delta(\xi_1)  =  \sqrt{2\pi}  \overline{\psi}
\delta(\xi_1).
\end{displaymath}
One has:
\begin{equation}
\left< \psi_D,\psi_D \right> = |\overline{\psi}|^2.
\label{Shvedov:e2a}
\end{equation}

\subsection{General case: algebraic approach}

Consider now  the  constrained  system with $n$ degrees of freedom and
$k$ linear constraints of a general form.  In the algebraic  approach,
states are  specified by wave functions $\psi_A(\xi)$ of $n$ variables
$\xi=(\xi_1,...,\xi_n)$. The constraints can be written as
\begin{equation}
\hat{\Lambda}_a = \Omega({\cal P}^{(a)},{\cal Q}^{(a)}) \equiv  \Omega
({\cal X}^{(a)}).
\label{Shvedov:e4}
\end{equation}
Here the linear combination of coordinate and momenta operators
\begin{displaymath}
\Omega(P,Q) =    \sum_{j=1}^n    (P_j\xi_j    -    Q_j     \frac{1}{i}
\frac{\partial}{\partial \xi_j})
\end{displaymath}
is an operator-valued 1-form satisfying the commutation relation
\begin{displaymath}
e^{i\Omega(X)} e^{i\Omega(X')}   =   e^{i\Omega(X+X')}  e^{\frac{i}{2}
\omega(X,X')}
\end{displaymath}
with
\begin{displaymath}
\omega(X,X') =
\omega\left(
\left( \begin{array}{c} P \\ Q \end{array} \right),
\left( \begin{array}{c} P' \\ Q' \end{array} \right)
\right) = \sum_j (P_jQ_j' - P_j'Q_j)
\end{displaymath}
being a  symplectic  2-form  of   classical   mechanics.   Then,   the
constraints ~\eqref{Shvedov:e4} commutes iff the isotropic condition is satisfied:
\begin{displaymath}
\omega ({\cal X}^{(a)},{\cal X}^{(b)}) = 0.
\end{displaymath}
Therefore, system  with  $k$  linear constraints can be specified by a
$k$-dimensional isotropic plane in $2n$-dimensional phase space:
\begin{displaymath}
{\cal L}_k = span \{ {\cal X}^{(1)},...,{\cal X}^{(k)}\}.
\end{displaymath}
The inner product in the algebraic approach can be presented as
\begin{equation}
\left< \psi_A^I, \psi_A^{II} \right> = \int_{{\cal L}_k} d\mu(X)
(\psi_A^I, e^{i\Omega(X)} \psi_A^{II})
\label{Shvedov:e5}
\end{equation}
with $d\mu(X)$ being a measure on  ${\cal  L}_k$  that is  invariant
under shifts. In the coordinate notations, inner product ~\eqref{Shvedov:e5} can be
rewritten as
\begin{displaymath}
\left< \psi_A^I, \psi_A^{II} \right> \sim
(\psi_A^I, \prod_a \delta (\hat{\Lambda}_a) \psi_A^{II}),
\end{displaymath}
this is in agreement with table 2.

Denote by  ${\cal  S}(\mathbb{R}^n,{\cal  L}_k)$  the  Shwartz  space  ${\cal
S}(\mathbb{R}^n)$ with the inner product ~\eqref{Shvedov:e5}. It is a pre-Hilbert space.

\begin{lemma} \label{Shvedov:l1} 1.  Let $\psi_A^I,  \psi_A^{II} \in {\cal S}(\mathbb{R}^n)$. Then
the integral ~\eqref{Shvedov:e5}  converges.  It  is  continuous  with  respect  to
$\psi_A^I$ and $\psi_A^{II}$ in ${\cal S}(\mathbb{R}^n)$-topology.
\\
2. For $\psi_A \in {\cal S}(\mathbb{R}^n)$, one has
$\left< \psi_A, \psi_A \right> \ge 0$.
\end{lemma}

It is important to note that for the case $\psi_A = \Omega(X) \varphi$
for some $X \in {\cal L}_k$,  one has $\left< \psi_A,\psi_A \right>  =
0$. Therefore,  there  are  zero-norm  states in the pre-Hilbert space
${\cal S}(\mathbb{R}^n,{\cal L}_k)$.  One should  set  $\psi_A  \sim  0$  iff
$\left<\psi_A,\psi_A\right> =  0$  and consider equivalence classes of
functions. Making use of the standard procedure of  factorization  and
completeness, one obtains the Hilbert space
$$
{\cal H} = \overline{{\cal S}(\mathbb{R}^n,{\cal L}_k)/\sim}.
$$
This is a Hilbert state space of the quantum theory.

Investigate examples  of observables --- operators acting in $\cal H$.
The simplest  example  of  observable  is  a  linear  combination   of
coordinate and   momenta   operators  ~\eqref{Shvedov:e4}.  However,  the  operator
$\Omega(Y)$ conserves the equivalence relation iff
$[\Omega(Y),\Omega(X)] =   0$   for   all   $X\in   {\cal   L}_k$,  or
$\omega(Y,{\cal L}_k)  =  0$.  This   means   that   $Y$   should   be
skew-orthogonal to    the   plane   ${\cal   L}_k$:   $Y   \in   {\cal
L}_k^{\perp\omega}$.

\begin{lemma} \label{Shvedov:l2}
1. Let $Y \in {\cal L}_k^{\perp\omega}$. Then the operator
$\Omega(Y):{\cal S}(\mathbb{R}^n,{\cal L}_k) \to
{\cal S}(\mathbb{R}^n,{\cal L}_k)$ is Hermitian and takes equivalent
states to equivalent.
\\
2. The operator $e^{i\Omega(Y)}$ for
$Y \in  {\cal  L}_k^{\perp\omega}$  is uniquely extended to an unitary
operator in $\cal H$.
\\
3. Let $Y_n \in {\cal L}_k^{\perp \omega}$ is  a  sequence  such  that
$Y_n \to_{n\to\infty} 0$. Then $e^{i\Omega(Y_n)} \to 1$ strongly.
\end{lemma}

\subsection{General case: Dirac approach}

Another construction of physical Hilbert space can be obtaned with the
help of the Dirac approach.  According to table 2, Dirac wave function
$\psi_D$ is related to $\psi_A$ as
\begin{equation}
\psi_D =  \int_{{\cal  L}_k} d\mu(X) e^{i\Omega(X)} \psi_A \equiv \eta
\psi_A.
\label{Shvedov:e6}
\end{equation}
It follows  from  lemma  ~\ref{Shvedov:l1}  that $\psi_D$ is a distribution from
${\cal S}'(\mathbb{R}^n)$.  Denote set of all distributions $\psi_D$  of  the
form ~\eqref{Shvedov:e6} as
\begin{displaymath}
\{ \psi_D  \}  \equiv  \check{\cal S} (\mathbb{R}^n,{\cal L}_k) \subset {\cal
S}'(\mathbb{R}^n).
\end{displaymath}

\begin{lemma} \label{Shvedov:l3}
$\psi_A \sim 0$ iff $\eta \psi_A = 0$.
\end{lemma}

Therefore, equivalent  wave  functions $\psi_A$ are mapped to the same
distribution $\psi_D$. This allows us to introduce the operator
\begin{displaymath}
\eta_0: {\cal S}(\mathbb{R}^n,{\cal L}_k)/\sim \to
\check{\cal S} (\mathbb{R}^n,{\cal L}_k)
\end{displaymath}
of the form
\begin{displaymath}
\eta_0[\psi_A] = \eta \psi_A, \quad \psi_A \in [\psi_A].
\end{displaymath}

\begin{lemma} \label{Shvedov:l4}
1. Let $\{[\psi_{A,n}]\}$ be a fundamental sequence from
${\cal S}(\mathbb{R}^n,{\cal L}_k)/\sim$.Then the sequence
$\psi_{D,n} = \eta_0[\psi_{A,n}]$ converges in ${\cal S}'(\mathbb{R}^n)$.
\\
2. Let $\eta_0[\psi_{A,n}] \to_{n\to\infty} 0$
in ${\cal S}'(\mathbb{R}^n)$. Then
$\left< \psi_{A,n},\psi_{A,n} \right> \to 0$.
\end{lemma}

Lemma ~\ref{Shvedov:l4}  means  that  the  operator  $\eta_0$  can  be  uniquely
extended to the operator $\overline{\eta}:  \overline{\psi} \in  {\cal
H} \mapsto \psi_D \equiv \overline{\eta} \overline{\psi} \in
{\cal S}'(\mathbb{R}^n)$.  The operator  $\overline{\eta}$  is  a  one-to-one
correspondence. It   maps  $\cal  H$  to  ${\cal  H}_D  \subset  {\cal
S}'(\mathbb{R}^n)$, which is a physical Hilbert space in the Dirac  approach.
The Dirac onnert product is
\begin{displaymath}
\left< \overline{\eta} \overline{\psi},
\overline{\eta} \overline{\psi}\right>_D =
\left< \overline{\psi},
\overline{\psi}\right>.
\end{displaymath}

\subsection{Explicit form of Dirac state space}

The considered  definition  of  space  ${\cal  H}_D$  is indirect.  It
requires the  algebraic  approach.  Let   us   present   an   explicit
description of space ${\cal H}_D$.

We say  that  $k$-dimensional  isotropic  plane ${\cal G}_k$ is a {\it
gauge surface} for ${\cal L}_k$ iff the  2-form  $\omega(\cdot,\cdot)$
is nondegenerate on ${\cal L}_k + {\cal G}_k$.

\begin{lemma}
\label{Shvedov:l5}
1. For  each isotropic plane ${\cal L}_k$ there exists a gauge surface
${\cal G}_k$.
\\
2. For any basis ${\cal X}^{(1)}$,...,${\cal X}^{(k)}$ on ${\cal L}_k$
there exists a basis
${\cal Y}^{(1)}$,...,${\cal Y}^{(k)}$ on ${\cal G}_k$ such that
\begin{displaymath}
\omega({\cal X}^{(a)},{\cal Y}^{(b)}) = \delta_{ab},
\end{displaymath}
3. The following decomposition takes place:
\begin{displaymath}
\mathbb{R}^{2n} =  {\cal  L}_k  +  {\cal  G}_k  +   ({\cal   L}_k   +   {\cal
G}_k)^{\perp\omega}.
\end{displaymath}
\end{lemma}

\begin{lemma}
\label{Shvedov:l6}
Let $\psi_D^I,  \psi_D^{II} \in {\cal S}'(\mathbb{R}^n)$.  Then expression of
the form
\begin{displaymath}
R(Y) = (\psi_D^I, e^{i\Omega(Y)} \psi_D^{II}), \quad Y \in \mathbb{R}^{2n}
\end{displaymath}
specifies a distribution from ${\cal S}'(\mathbb{R}^{2n})$.  It is continuous
with respect to $\psi_D^I$ and $\psi_D^{II}$.
\end{lemma}

\begin{lemma}
\label{Shvedov:l7}
Distribution $\psi_D \in {\cal S}'(\mathbb{R}^n)$ is  of  ${\cal  H}_D$-class
iff two conditions are satisfied:
\begin{itemize}
\item $\Omega(X) \psi_D = 0$ for all $X \in {\cal L}_k$;
\item distribution  $(\psi_D,e^{i\Omega(Y)}   \psi_D)$   possesses   a
restriction on ${\cal G}_k$.
\end{itemize}
\end{lemma}

Note that   distribution   $\Phi(x,y)$,
$x\in \mathbb{R}^{n_1}$, $y\in \mathbb{R}^{n_2}$
possesses  a  restriction  on
$y=0$-plane iff the function
\begin{displaymath}
\Phi_{\varphi_1}(y) = \int dx \varphi_1(x) \Phi(x,y).
\quad
\varphi_1 \in {\cal S}(\mathbb{R}^{n_1})
\end{displaymath}
is continuous with respect to $y$ in the vicinity of the point $y=0$.

Let us obtain the expression for  the  inner  product  of  Dirac  wave
functions.

\begin{lemma}
\label{Shvedov:l9a}
1. Let   ${\cal  L}_k$  be  isotropic  plane  with  invariant  measure
$d\mu(X)$, ${\cal G}_k$ be a  gauge  surface  for  ${\cal  L}_k$  with
invariant measure $d\sigma(Y)$. Then
\begin{equation}
\int d\mu(X) \int d\sigma(Y) \rho(Y)
e^{i\omega(X,Y)} = \rho(0) \cdot \Delta, \quad \Delta = const.
\label{Shvedov:e7a}
\end{equation}
2. Let $\rho(Y) \in {\cal S}({\cal G}_k)$ and $\rho(0) = 1/\Delta$.
Then
\begin{equation}
\left< \psi_D,\psi_D\right>_D = \int_{{\cal G}_k} d\sigma(Y) \rho(Y)
(\psi_D, e^{i\Omega(Y)} \psi_D).
\label{Shvedov:e7}
\end{equation}
\end{lemma}

Let us illustrate formula ~\eqref{Shvedov:e7}, making use of simple one-dimansional
examples.

1. Let $\hat{\Lambda}_1 = \hat{p}_1$.  Then the Dirac  wave  fucntions
$\psi_D=const$. Surfaces ${\cal L}_1$ and ${\cal G}_1$ are axis:
\begin{displaymath}
{\cal L}_1 = \{ (p_1,q_1) | p_1 = 0\},
\quad
{\cal G}_1 = \{ (p_1,q_1) | q_1 = 0\}
\end{displaymath}
with measures $dp_1$ and $dq_1$. Formula ~\eqref{Shvedov:e7a} is rewritten as
\begin{displaymath}
\int dq_1 dp_1 \rho(p_1) e^{ip_1q_1} = \rho(0) \cdot \Delta,
\end{displaymath}
so that $\Delta = 2\pi$. The inner product ~\eqref{Shvedov:e7} is
\begin{displaymath}
\left<\psi_D,\psi_D \right>_D =
\int dp_1 \rho(p_1) \int dq_1 \psi_D^*(q_1)
e^{ip_1\hat{q}_1} \psi_D(q_1) = 2\pi \rho(0) |\psi_D|^2.
\end{displaymath}
This is in agreement with relation ~\eqref{Shvedov:e2}.

2. Let $\hat{\Lambda}_1 = \hat{q}_1$.  Then the Dirac  wave  functions
are of the form
$\psi_D(\xi) = \sqrt{2\pi} \overline{\psi} \delta(\xi_1)$.
${\cal L}_1$ and ${\cal G}_1$ are axis
\begin{displaymath}
{\cal L}_1 = \{ (p_1,q_1) | q_1 = 0\}.
\quad
{\cal G}_1 = \{ (p_1,q_1) | p_1 = 0\},
\end{displaymath}
Formula ~\eqref{Shvedov:e7a} is rewritten as
\begin{displaymath}
\int dp_1 dq_1 \rho(q_1) e^{-ip_1q_1} = \rho(0) \cdot \Delta,
\end{displaymath}
so that $\Delta = 2\pi$. the inner product ~\eqref{Shvedov:e7} is
\begin{displaymath}
\left<
\psi_D,\psi_D \right>_D =
\int dq_1 \rho(q_1) \int d\xi_1 \psi_D^*(\xi_1)
e^{-iq_1 \frac{1}{i} \frac{\partial}{\partial \xi_1}} \psi_D(\xi_1) =
2\pi |\overline{\psi}|^2 \rho(0);
\end{displaymath}
this agrees with ~\eqref{Shvedov:e2a}.

\section{Properties of Gaussian and quasi-Gaussian states}

Gaussian and  quasi-Gaussian  states  play  an   important   role   in
investigation of quadratic Hamiltonians. A Gaussian wave function is
\begin{displaymath}
\psi_A(\xi) = const \exp \left\{
\frac{i}{2} \sum_{jk} \xi_j A_{jk} \xi_k\right\}
\end{displaymath}
with complex matrix $A$. A quasi-Gaussian wave function is of the form
of a product of a polynomial $P(\xi)$ by a Gaussian functon:
\begin{displaymath}
\psi_A(\xi) = const P(\xi) \exp \left\{
\frac{i}{2} \sum_{jk} \xi_j A_{jk} \xi_k\right\}
\end{displaymath}

An important  property  of Gaussian and quasi-Gaussian states is their
invariance under  evolution  for  the  quadratic   Hamiltonian   case.
Gaussian states  evolve  to Gaussian,  quasi-Gaussian states evolve to
quasi-Gaussian.

It is also important to note that  set  of  quasi-Gaussian  states  is
complete; this  means that arbitrary wave function can be approximated
by a  quasi-Gaussian  function.   If   evolution   transformation   is
calculated for the quasi-Gaussian initial consitions, one can uniquely
extend it to the general case.

The Maslov    complex     germ     \cite{Shvedov:Maslov2}     $r(A)$     is     a
representtion-invariant characteristics    of   Gaussian   state.   By
definition,
\begin{displaymath}
r(A) = \{ Y | \Omega(Y) \psi_A = 0 \}
\end{displaymath}
is a subspace of complexified phase space $\mathbb{C}^{2n}$. It is known that
for systems without constraints $r(A)$ is a graph of the operator with
matrix $A$:
\begin{displaymath}
r(A) = \{ (P=AQ,Q)|Q \in \mathbb{C}^n\}.
\end{displaymath}
The following  properties  of  the  Maslov  complex germ are satisfied
(cf.\cite{Shvedov:Maslov2}):
\begin{gather}
\mbox{isotropic property: $Y_1,Y_2 \in r$ $\Rightarrow$ $\omega(Y_1,Y_2) = 0$};
\nonumber
\\
\mbox{positiveness: $Y   \in    r$,    $Y\ne    0$    $\Rightarrow$
$\frac{1}{i} \omega(Y,Y^*) > 0$}.
\label{Shvedov:e8}
\end{gather}
An inverse  property  is  also  satisfied.   Namely,   introduce   the
projection operators on the coordinate and momenta planes:
\begin{displaymath}
B: r(A) \to \mathbb{C}^n, \quad
C: r(A) \to \mathbb{C}^n
\end{displaymath}
of the form
\begin{displaymath}
Y = (P,Q) \in r(A) \mapsto BY = P, CY=Q.
\end{displaymath}
Then for any $n$-dimensional surface $r \subset  \mathbb{C}^{2n}$  satisfying
relations ~\eqref{Shvedov:e8}, the following properties take place:
\begin{displaymath}
\mathbb{C}^{2n} = r + r^*, \quad r \cap r^* = \{0\},
\quad
\mbox { $C$ is invertible, $r=r(A)$ for $A=BC^{-1}$. }
\end{displaymath}

For constrained systems,  property $\Omega(Y)\psi_A = 0$  entering  to
definition of  the  Maslov  complex  germ  may  be viewed in
${\cal S}(\mathbb{R}^n)$ and  in  $\cal  H$.  Thus,  two  different  objects,
"S-germ" and "H-germ" should be introduced:
\begin{gather*}
r(A) = \{ Y| \Omega(Y) \psi_A = 0 \mbox{ in } {\cal S}(\mathbb{R}^n)\} \mbox{
- S-germ;}
\\
\check{r}(A) =
\{ Y| ||\Omega(Y) \psi_A|| = 0 \} \mbox{
- H-germ.}
\end{gather*}
Note that S-germ is a graph of $A$, while H-germ has another form.

Introduce the following notations for the subspaces of $r(A)$:
\begin{gather*}
r_{\perp}(A) = r(A) \cap ({\cal L}_k^{\mathbb{C}})^{\perp\omega},
\\
r_-(A) = \{ Y \in  r(A)|\omega(Y,Y')  =  0  \mbox{  for  all  $Y'  \in
r_{\perp}(A)$ } \}.
\end{gather*}

\begin{lemma}
\label{Shvedov:l8}
1. Any element $X \in {\cal L}_k$ can be uniquely decomposed as  $X  =
X_- + X_-^*$, $X_- \in r_-(A)$.
\\
2. $dim r_{\perp}(A) = n-k$, $dim r_-(A) = k$.
\\
3. The operator $P_-: {\cal L}_k^{\mathbb{C}} \to r_-(A)$ of the form $P_-: X
\mapsto X_-$ is a linear one-to-pne map.
\end{lemma}

For any  linear one-to-one map of measure spaces $P:{\cal L} \to {\cal
L}'$, introduce  the  notion  of  Jacobian.   Let   $x_1,...,x_k$   be
coordinates on $\cal L$, $x_1',...,x_k'$ be coordinates on ${\cal L}'$,
\begin{displaymath}
d\mu = J dx_1...dx_k,
\quad
d\mu' = J' dx_1' ... dx_k'
\end{displaymath}
be measures  on  $\cal L$ and ${\cal L}'$,  $P_{ij}$ be matrix of $P$.
Then denote
\begin{displaymath}
\Delta(P) \equiv |det P_{ij}| |J'|/|J|.
\end{displaymath}

\begin{lemma} \label{Shvedov:l9}
The following relation is satisfied for the inner product:
\begin{displaymath}
\left<\psi_A,\psi_A \right> =
\int_{{\cal L}_k} d\mu(X)
e^{\frac{i}{2} \omega(P_-X,(P_-X)^*)} (\psi_A,\psi_A) =
(2\pi)^{\frac{k_n}{2}} |c|^2 \frac{\Delta (C)}{\Delta (P_-)}.
\end{displaymath}
\end{lemma}

Let us generalize properties ~\eqref{Shvedov:e8} to the constrained systems.

\begin{lemma}
\label{Shvedov:l10}
The set $\check{r}(A) \equiv \check{r}$ is an $n$-dimensional subspace
of $\mathbb{C}^{2n}$. It has the form
\begin{displaymath}
\check{r}(A) = r_{\perp}(A) + {\cal L}_k^{\mathbb{C}},
\quad
r_{\perp}(A) \cap {\cal L}_k^{\mathbb{C}} = \{ 0\}
\end{displaymath}
and satisfies the following properties:
\begin{gather}
Y_1,Y_2 \in \check{r} \mbox{ $\Rightarrow$ } \omega(Y_1,Y_2) = 0,
\nonumber \\
Y \in {\cal L}_k^{\mathbb{C}} \mbox{ $\Rightarrow$ } Y\in \check{r}, \omega(Y,Y^*) = 0,
\label{Shvedov:e9}
\\
Y \in \check{r} \setminus {\cal L}_k^{\mathbb{C}} \mbox{ $\Rightarrow$ }
\frac{1}{i} \omega(Y,Y^*) > 0.
\nonumber
\end{gather}
\end{lemma}

An inverse statement is also correct.

\begin{lemma} \label{Shvedov:l11}
Let $n$-dimensional  subspace  $\check{r}   \subset   \mathbb{C}^{2n}$   obey
relations ~\eqref{Shvedov:e9}. Then
\begin{displaymath}
({\cal L}_k^{\mathbb{C}})^{\perp\omega} = \check{r} + \check{r}^*,
\quad
\check{r} \cap \check{r}^* = {\cal L}_k^{\mathbb{C}}
\end{displaymath}
and $\check{r} = \check{r}(A)$ for any $A$.
\end{lemma}

A remarkable property of H-germ is as follows:  equivalent  Gaussian
states correspond to the same complex H-germ.

\begin{lemma} \label{Shvedov:l12}
Let
\begin{displaymath}
\psi_A^I = c^I e^{\frac{i}{2} \sum_{jk} \xi_j A_{jk}^I \xi_k},
\quad
\psi_A^{II} = c^{II}
e^{\frac{i}{2} \sum_{jk} \xi_j A_{jk}^{II} \xi_k}
\end{displaymath}
be two Gaussian states. Then $\check{r} (A^I) = \check{r}(A^{II})$ iff
$\psi_A^I \sim c \psi_A^{II}$ for some complex number $c$.
\end{lemma}

Consider the Dirac wave function $\psi_D$  corresponding  to  Gaussian
state. An  explicit  form of $\psi_D$ is also Gaussian,  provided that
${\cal L}_k$ is uniquely projectable to the $P=0$-plane:
\begin{displaymath}
\psi_D(\xi) = \check{c}
e^{\frac{i}{2} \sum_{jk} \xi_j \check{A}_{jk} \xi_k}.
\end{displaymath}
To write explicit forms of $\check{c}$ and $\check{A}$,  introduce the
following notations. By
\begin{displaymath}
\check{B}: \check{r}(A) \to \mathbb{C}^n,
\quad
\check{C}: \check{r}(A) \to \mathbb{C}^n
\end{displaymath}
we denote  projectors  to  momenta  and  coordinate planes.  Let $\Pi:
\check{r}(A) \to {\cal L}_k^{\mathbb{C}}$  be  the  operator  taking  $Y  \in
\check{r}(A)$ to  $\Pi  Y  \in  {\cal  L}_k^{\mathbb{C}}$  obtained  from the
relation
\begin{displaymath}
Y = Y_{\perp} + \Pi Y, \quad Y_{\perp} \in r_{\perp}(A).
\end{displaymath}
Introduce also the operator ${\cal P}_-:\check{r}(A) \to r(A)$ of  the
form:
\begin{displaymath}
{\cal P}_-Y = \left\{
\begin{array}{c}
Y, \quad Y \in r_{\perp}(A), \\
P_-Y, \quad Y \in {\cal L}_k^{\mathbb{C}}.
\end{array}
\right.
\end{displaymath}

\begin{lemma} \label{Shvedov:l13}
The following relations are satisfied:
\begin{displaymath}
\check{A} = \check{B} \check{C}^{-1},
\quad
\check{c} = \check{c}
\int_{{\cal L}_k} d\mu(X) e^{-\frac{i}{2} \omega(X,C^{-1} \check{C}X)}
= c (2\pi)^{k/2}
\frac{\sqrt{det (\Pi \check{C}^{-1} C {\cal P}_-)}}{\Delta(P_-)}.
\end{displaymath}
\end{lemma}

For the simplest one-dimensional example, with constraint
\begin{displaymath}
\hat{\Lambda}_1 =    {\cal    P}\xi    -    {\cal    Q}    \frac{1}{i}
\frac{\partial}{\partial \xi},
\end{displaymath}
the Dirac wave function has the form
\begin{displaymath}
\psi_D(\xi) = \left\{
\begin{array}{c}
const e^{\frac{i}{2} \frac{\cal P}{\cal Q} \xi^2}, \quad Q \ne 0,
\\
const \delta(\xi), \quad Q = 0.
\end{array}
\right.
\end{displaymath}

The following  lemma tells us that set of all quasi-Gaussian states is
dense.

\begin{lemma} \label{Shvedov:l14}
1. Set of all linear combinations of vectors of the form
$e^{i\Omega(Y)} [\psi_A]$, $Y \in {\cal L}_k^{\perp\omega}$,
where $[\psi_A]$ is a Gaussian vector, is dense in $\cal H$.
\\
2. Set of all linear combinations of vectors of the form
$\Omega(Y_1^*) ... \Omega(Y_p^*)
[\psi_A]$, $Y_1^*,...,Y_p^* \in \check{r}^*(A)$,
where $[\psi_A]$ is a Gaussian vector, is dense in $\cal H$.
\end{lemma}

\section{Quadratic Hamiltonians}

let us  investigate  the  properties of quadratic Hamiltonians for the
constrained systems.  To  present  the  results  in  invariant   form,
introduce the following notations.

Let ${\cal M} = \mathbb{R}^{2n}$ be classical phase space. To element
\begin{displaymath}
\Gamma = \frac{1}{2} \sum_{ij=1}^{2n}
\Gamma_{ij} Z^{(i)} \otimes Z^{(j)} \in Sym {\cal M} \otimes {\cal M}
\end{displaymath}
we assign the  operator
\begin{displaymath}
\Omega_2(\Gamma) =
\frac{1}{2} \sum_{ij=1}^{2n} \Gamma_{ij}
\Omega(Z^{(i)})\Omega(Z^{(j)}).
\end{displaymath}
A quadratic Hamiltonian has the form
\begin{displaymath}
H = \Omega_2(\Gamma) + {\varepsilon},  \Gamma \in Sym {\cal M} \otimes
{\cal M},
\end{displaymath}
${\varepsilon}$ is an operator of multiplication by a real number.

\begin{lemma} \label{Shvedov:l15}
1. $H$ conserve the equivalence relation
($f_1 \sim f_2 \Rightarrow Hf_1 \sim Hf_2$) iff
$\Gamma \in    Sym    {\cal    L}_k^{\perp\omega}    \otimes     {\cal
L}_k^{\perp\omega} + Sym {\cal L}_k \otimes {\cal M}$.
\\
2. For any quadratic Hamiltonian $H$ there exists
$\Gamma' \in    Sym    {\cal    L}_k^{\perp\omega}    \otimes    {\cal
L}_k^{\perp\omega}$ such that
\begin{displaymath}
Hf \sim [\Omega_2(\Gamma') + {\varepsilon}'] f.
\end{displaymath}
\end{lemma}

Therefore, without  loss  of  generality one can set $\Gamma \in {\cal
L}_k^{\perp\omega} \otimes {\cal L}_k^{\perp\omega}$.

\begin{lemma} \label{Shvedov:l16}
1. $H$ is a symetric operator on the set of quasi-Gaussian vectors.
\\
2. Quasi-Gaussian vectors are analytic for the operator $H$.
\end{lemma}

It is convenent to introduce a reduced classical space
\begin{displaymath}
{\cal R} = {\cal L}_k^{\perp\omega}/{\cal L}_k.
\end{displaymath}
Since $[\Omega(Y)f]=0$ for $Y \in {\cal L}_k$, the operator
$\overline{\Omega}(\overline{Y}) =   \Omega(Y)$  is  well-defined  for
$\overline{Y} \in {\cal R}$.  Analogously,  one defines  the  operator
$\overline{\Omega}_2(\overline{\Gamma})$ for   $\overline{\Gamma}  \in
Sym {\cal R} \otimes {\cal R}$.  The evolution equation can be written
as
\begin{equation}
i \frac{d\psi_A}{dt} \sim
[\overline{\Omega}_2(\overline{\Gamma}) + {\varepsilon}]\psi_A.
\label{Shvedov:e10}
\end{equation}
It is known \cite{Shvedov:MSh} that relation
\begin{equation}
\left[
i \frac{d}{dt}     -     \overline{\Omega}_2(\overline{\Gamma})      -
{\varepsilon}, \overline{\Omega} (\overline{Y}(t))
\right] = 0
\label{Shvedov:e11}
\end{equation}
is satisifed  iff  $\overline{Y}(t)$  satisfies  the reduced classical
Hamiltonian system:
\begin{equation}
\frac{d\overline{Y}}{dt} = \overline{Y} \circ \overline{\Gamma}.
\label{Shvedov:e12}
\end{equation}
The $\circ$-product    of    $\overline{Y}    \in    {\cal   R}$   and
$\overline{\Gamma} \in Sym {\cal R} \otimes {\cal R}$ is viewed in the
folloeing sense:
\begin{displaymath}
\overline{Y} \circ
\frac{1}{2} \sum_{ij}     \Gamma_{ij}      \overline{Z}_i      \otimes
\overline{Z}_j \equiv               \sum_{ij}              \Gamma_{ij}
\omega(\overline{Y},\overline{Z}_i) \overline{Z}_j \in {\cal R}.
\end{displaymath}
It follows from relation ~\eqref{Shvedov:e11} that the operator  $\overline{\Omega}
(\overline{Y}(t))$ transforms   solutions  of  evolution  equation  to
solution.

By $u_t:  \overline{Y}(0) \to \overline{Y}(t)$ we denote the classical
evolution transformation  taking  initial  condition  for  the  system
~\eqref{Shvedov:e12} to the solution of the Cauchy problem.

It happens that time-dependent Gaussian state
\begin{displaymath}
\psi_A(\xi,t) = c(t) e^{\frac{i}{2} \sum_{jk} \xi_j A_{jk}(t) \xi_k}
\end{displaymath}
satisfies evolution equation ~\eqref{Shvedov:e10} iff
\begin{gather*}
\check{r}(A(t)) = u_t \check{r}(A(0)),
\\
c(t) = c(0) \frac{\Delta({\cal P}_-(t))}{\Delta({\cal P}_-(0))}
\frac{1}{\sqrt{det (C(t) {\cal P}_-(t) u_t {\cal P}_-^{-1}(0) C(0))}}
= \frac{\check{c}(0)}{\sqrt{det (\check{C}(t) u_t  \check{C}^{-1}(0))}
}.
\end{gather*}

\section{Diagonalization of  quadratic  Hamiltonians   and   classical
stability}

The following   lemma  is  a  generalization  of  the  maslov  theorem
\cite{Shvedov:Maslov2} to the case of constrained syatems.

\begin{lemma} \label{Shvedov:l17}
1. Let   $\overline{\Omega}_2(\overline{\Gamma})$   have   a  Gaussian
eigenvector, Then Hamiltonian system ~\eqref{Shvedov:e12} for $\overline{Y}(t)  \in
{\cal R}$ is stable
\\
2. Let Hamiltonian system ~\eqref{Shvedov:e12} be stable in $\cal R$. Then:
\begin{itemize}
\item[(a)] there exist $2n-2k$ independent solutions of ~\eqref{Shvedov:e12}:
\begin{gather*}
\overline{Y}^{(I)}(t) = \overline{Y}^{(I)} e^{i\beta_It},
\quad
\overline{Y}^{(I)*}(t) = \overline{Y}^{(I)*} e^{-i\beta_It},
\quad
I = \overline{1,n-k},
\\
\frac{1}{i} \omega(\overline{Y}^{(I)},\overline{Y}^{(J)*}) =
\delta_{IJ},
\quad
\omega(\overline{Y}^{(I)},\overline{Y}^{(J)}) = 0;
\end{gather*}
\item[(b)] set    $\check{r}    =    span\{Y^{(1)},...,Y^{(n-k)},{\cal
L}_k^{\mathbb{C}}\}$ is a Maslov H-germ:  $\check{r} = \check{r}(A)$ for some
$A$;
\item[(c)] there   is   a   complete   set   of   eigenfunctions    of
$\overline{\Omega}_2(\overline{\Gamma})$:
\begin{displaymath}
\overline{\Omega}(\overline{Y}^{(1)*})^{N_1} ...
\overline{\Omega}(\overline{Y}^{(n-k)*})^{N_{n-k}} [\psi_A],
\end{displaymath}
with Gaussian eigenfunction $\psi_A$. The eigenvalues are:
\begin{displaymath}
\beta_1 (N_1+1/2) + ... + \beta_{n-k} (N_{n-k} + 1/2).
\end{displaymath}
\end{itemize}
\end{lemma}

It is important to note that stability is considered  in  the  reduced
classical phase space $\cal R$, not in $\cal M$.

\section{Specific featurs of infinite-dimensional case}

Quantum field models are usually infinite dimensional.  Therefore,  it
is important to generalize the main  notions  and  statements  to  the
infinite-dimensional case. The main notions are presented in table 3.

\begin{table}
\caption{Finite dimensional  systems  and  their  infinite-dimensional
analogs}
\begin{tabular}{|p{2.5cm}|p{12.5cm}|}
\hline
Finite-dimensional case
& Infinite-dimensional analogs
\\
\hline
wave function $\psi(\xi_1,...,\xi_n)$
& element
$\Psi = \left(
\begin{array}{c}
\Psi_0 \in \mathbb{C}
\\
\Psi_1 \in {\cal H}
\\
...
\\
\Psi_n \in Sym {\cal H}^{\otimes n}
\\
...
\end{array}
\right)$
of the  Fock  space  ${\cal  F}({\cal  H}) = \oplus_{n=0}^{\infty} Sym
{\cal H}^{\otimes n}$.
\\ \hline
coordinate and momenta operators $\hat{p}_i,\hat{q}_i$
& creation and annihilation operators $a^{\pm}[f]$, $f\in {\cal H}$,
$a^+[f] Sym f_1 \otimes ... \otimes f_k = \sqrt{k+1}
Sym f  \otimes  f_1 \otimes ...  \otimes f_k$,  $a^-[f] = (a^+[f])^+$,
$[a^-[f],a^+[g]] = (f,g)$.
\\ \hline
Schwartz space ${\cal S}(\mathbb{R}^n)$
&
$S \in {\cal F}({\cal H})$ --- set of all $\Psi$ such that
$||\Psi||_m = \max_n n^m ||\Psi_n|| < \infty$.
\\ \hline
Classical phase space
${\cal M} = \mathbb{R}^{2n} =
\{(P,Q)|P,Q\in \mathbb{R}^n\}$
& ${\cal H}_{\mathbb{R}}$ --- set $\cal H$ with  multiplication  on  real  nubers
only, real Hilbert space; $(f,g)_{\mathbb{R}} = Re (f,g)$.
\\ \hline
$\Omega(P,Q)$, $\omega$
&
$\Omega[f] = - i(a^+[f]-a^-[f])$, $\omega(f,g) = i ((f,g) - (g,f))$,
$[\Omega[f],\Omega[g]] = - i \omega(f,g)$.
\\ \hline
Pre-Hilbert space ${\cal S}(\mathbb{R}^n,{\cal L}_k)$
& $S$ with inner product
\begin{displaymath}
\left< \Psi^I,\Psi^{II}\right> =
\int_{{\cal L}_k} d\mu(\varphi)
(\Psi^I, e^{i\Omega[\varphi]} \Psi^{II}),
\quad \Psi^I, \Psi^{II} \in S,
\end{displaymath}
$d\mu(\varphi)$ --- invariant  measure  on  $k$-dimensional  isotropic
plane ${\cal L}_k$ in ${\cal H}_{\mathbb{R}}$.
\\ \hline
\end{tabular}
\end{table}

\begin{lemma} \label{Shvedov:l1f} (analog of lemma ~\ref{Shvedov:l1})
1. Integral
$\int_{{\cal L}_k} d\mu(\varphi)
(\Psi^I, e^{i\Omega[\varphi]} \Psi^{II})$ converges,  it is continuous
with respect to $\Psi^I,\Psi^{II}$ in $S$-topology.
\\
2. $\left< \Psi,\Psi\right> >0$.
\\
3. $\varphi \in {\cal L}_k \Rightarrow
\left< \Psi^I, \Omega[\varphi] \Psi^{II}\right> = 0$.
\end{lemma}

Analogously to finite-dimensional  case,  we  introduce  the  physical
Hilbert space
\begin{displaymath}
{\cal H} = \overline{S/\sim}.
\end{displaymath}

\begin{lemma} \label{Shvedov:l3f} (analog of lemma ~\ref{Shvedov:l2}) Let $Y \in {\cal H}_{\mathbb{R}}$  and
$\omega(Y,{\cal L}_k) = 0$.  Then $\Omega(Y)$ is Hermitian on $S$;  it
conserves the  equivalence  property;  $e^{i\Omega(Y)}$  is   uniquely
extended to an unitary operator in $\cal H$.
\end{lemma}

For the  infinite-dimensional case,  Gaussian states are introduced as
follows. The Fock vector of the form
\begin{displaymath}
\left(
\begin{array}{c}
1 \\ 0 \\...
\end{array}
\right) \equiv |0>
\end{displaymath}
is called as a vacuum vector. A Caussian state  is
\begin{displaymath}
\Psi_M = \exp\left[\frac{1}{2} a^+Ma^+ \right] |0>,
\end{displaymath}
where the quadratic form entering to the exponent is viewed as:
\begin{equation}
a^+Ma^+ \equiv \sum_{ij} a^+[f_i] (f_i^*,Mf_j) a^+[f_j],
\label{Shvedov:e14f}
\end{equation}
for any orthonormal basis $f_i$ in $\cal H$.  It is known  \cite{Shvedov:Berezin}
that expression  $\Psi_M$  specifies  a  vector  from $S$ iff $M$ is a
Hilbert-Schmidt operator and $||M|| < 1$.

Consider an analog of the complexified phase space  ${\cal  M}^{\mathbb{C}}$.
It should have the form
\begin{displaymath}
{\cal H}_{\mathbb{R}\mathbb{C}} \equiv {\cal H}_{\mathbb{R}} \otimes_{\mathbb{R}} \mathbb{C}.
\end{displaymath}
It happens that this space is isomorphic to
\begin{displaymath}
{\cal H}_{\mathbb{R}\mathbb{C}} \simeq
{\cal H} \oplus {\cal H}^*.
\end{displaymath}
The isomorphism is
\begin{displaymath}
f' \otimes_{\mathbb{R}} 1 + f'{}' \otimes_{\mathbb{R}} i \leftrightarrow
\left(
\begin{array}{c}
f' + i f'{}' = f \in {\cal H}
\\
f^{\prime *} + i f^{\prime\prime *} = \varphi^* \in {\cal H}^*
\end{array}
\right)
\end{displaymath}
In new notations:
\begin{gather*}
\left(
\begin{array}{c} f \\ \varphi^* \end{array}
\right)^* =
\left(
\begin{array}{c} \varphi \\ f^* \end{array}
\right),
\quad
\left(
\left(
\begin{array}{c} f \\ \varphi^* \end{array}
\right),
\left(
\begin{array}{c} g \\ \chi^* \end{array}
\right)
\right) =
\frac{1}{2} [(f,g) + (\varphi^*,\chi^*)];
\\
\Omega\left[\left(
\begin{array}{c} f \\ \varphi^* \end{array}
\right) \right] =
- i (a^+[f]- a^-[\varphi]),
\quad
\left<
\left(
\begin{array}{c} f \\ \varphi^* \end{array}
\right),
\left(
\begin{array}{c} g \\ \chi^* \end{array}
\right)
\right> =
i [\varphi^*[g] - \chi^*[f]].
\end{gather*}

A Maslov complex germ is defined as follows.

S-germ is set of all $\chi \in {\cal H} \oplus {\cal H}^*$ such that
$\Omega[\chi]\Psi_M = 0$ is $S$:
\begin{displaymath}
r(M) = \left\{
\left(
M\varphi^* \\ \varphi^*
\right), \varphi^* \in {\cal H}^*
\right\}
\end{displaymath}

H-germ $\check{r}(M)$ is a set of all $\chi \in {\cal H} \oplus  {\cal
H}^*$ such that $\omega(\chi,{\cal L}_k) = 0$ and $\Omega[\chi] \Psi_M
\sim 0$ in $\cal H$.

\begin{lemma} \label{Shvedov:l11f} (analog of ~\ref{Shvedov:l10}, ~\ref{Shvedov:l11}, ~\ref{Shvedov:l12})
1.  $\check{r}(M)  =  r_{\perp}(M)
\oplus {\cal L}_k^{\mathbb{C}}$.
\\
2. The  plane  $\check{r}  =  \check{r}(M)$  satisfies  the  following
properties:
\begin{itemize}
\item[(a)]
$\chi \in {\cal L}_k^{\mathbb{C}} \Rightarrow \chi \in \check{r},
\frac{1}{i} \omega(\chi,\chi^*) > 0$.
\item[(b)] $\chi   \in   \check{r},   \chi\notin   {\cal    L}_k^{\mathbb{C}}
\Rightarrow \frac{1}{i} \omega(\chi,\chi^*) > 0$.
\item[(c)]
$\chi_1,\chi_2 \in \check{r} \Rightarrow \omega(\chi_1,\chi_2) = 0$.
\item[(d)]
Let $\chi \in {\cal H} \oplus {\cal H}^*,
\omega(\chi,{\cal L}_k^{\mathbb{C}}) = 0$. Then
$\chi=\chi_+ +   \chi_-$,   $\chi_-   \in   \check{r}$,   $\chi_+  \in
\check{r}$. Vectors $\chi_{\pm}$ are determined up to adding  elements
form ${\cal L}_k^{\mathbb{C}}$, and the operators
\begin{displaymath}
{\cal P}: {\cal L}_k^{\mathbb{C}\perp\omega} \to \check{r}/{\cal L}_k^{\mathbb{C}},
\quad
{\cal P}^*:    {\cal   L}_k^{\mathbb{C}\perp\omega}   \to   \check{r}^*/{\cal
L}_k^{\mathbb{C}}
\end{displaymath}
of the form
\begin{displaymath}
{\cal P} \chi = [\chi_-], \quad
{\cal P}^* \chi = [\chi_+].
\end{displaymath}
are well-defined.
\item[(e)]
${\cal P}|_{{\cal H} =
\left\{ \left(
\begin{array}{c} f \\ 0 \end{array}
\right) \right\}
}$ is a Hilbert-Schmidt operator.
\end{itemize}
3. Let  $\check{r}  \subset  {\cal  L}_k^{\mathbb{C}  \perp  \omega}$ satisfy
properties (a)-(e). Then $\check{r} = \check{r}(M)$ for some $M$.
\\
4. $\check{r}(M_1)  =  \check{r}(M_2)$  iff  $\Psi_{M_1}  \sim   const
\Psi_{M_2}$.
\end{lemma}

The infinite-dimensional quadratic Hamiltonians are of the form
\begin{displaymath}
H =  \frac{1}{2}  a^+Aa^+  +  a^+Ba^-  +   \frac{1}{2}   a^-A^*a^-   +
{\varepsilon},
\end{displaymath}
where ${\varepsilon}$    is    a    multiplicator   by
a complex number
${\varepsilon}$, definition of quadratic forms are analogous to
~\eqref{Shvedov:e14f}. Here ${\cal L}_k \in D({B})$,  while  the  kernel  of  the
operator $A$ belongs to $D(B) \otimes D(B)$.

Classical Hamiltonian system is of the form
$\dot{\chi} = L\chi$ with
\begin{displaymath}
L \left(
\begin{array}{c}
\varphi \\ \varphi^*
\end{array}
\right) =
\left(
\begin{array}{c}
-iB\varphi - iA\varphi^* \\
iA^* \varphi + i (B\varphi)^*
\end{array}
\right)
\end{displaymath}
Impose the conditions:
\begin{displaymath}
L {\cal L}_k \subset {\cal L}_k,
\quad
Im {\varepsilon} = - \frac{1}{2} Tr L|_{{\cal L}_k}.
\end{displaymath}

\begin{lemma} \label{Shvedov:l16f} (analog of lemma ~\ref{Shvedov:l16})
1. $H$ is essentially self-adjoint on $S \cap D(a^+Ba^-)$.
For the case $M \in D(B) \otimes D(B)$, $\chi_1,...,\chi_p \in D(B)$,
one has $\Omega[\chi_1]... \Omega[chi_p] \Psi_M \in D(H)$.
\\
2. Let
\begin{equation}
\frac{d\chi}{dt}   =   L^{\mathbb{C}}   \chi.
\label{Shvedov:e21}
\end{equation}
Then  the  operator
$\Omega[\chi(t)]$ commutes with $i\frac{d}{dt} - H$.
\\
3. Let $\varphi_0,f_0 \in D(B)$.  Then there exists a unique  solution
$\chi(t) =
\left(
\begin{array}{c}
f(t) \\ \varphi^*(t)
\end{array}
\right)$ to  the  Caushy  problem  for equation ~\eqref{Shvedov:e21}.  The operator
$u_t: \chi_0 \mapsto \chi(t)$ is bounded;  it is uniquely extended  to
${\cal H} \oplus {\cal H}^*$.  The operator $u_t$ conserves the 2-form
$\omega$.

Consider the reduced classical space
\begin{displaymath}
{\cal R} = {\cal L}_k^{\perp\omega}/{\cal L}_k,
\end{displaymath}
with the norm
\begin{displaymath}
||[\chi]||_{\cal R} = inf_{\chi \in [\chi]} ||\chi||.
\end{displaymath}
The operator $u_t$ can be considered on this factorspace.
\end{lemma}

\begin{lemma} \label{Shvedov:l18f} (analog of lemma ~\ref{Shvedov:l17})
Let $H$  have  a Gaussian eigenfunction and $M \in D(B) \otimes D(B)$.
Then classical system
$\frac{d\chi}{dt} = L\chi$ is stable in $\cal R$-topology.
\end{lemma}

\subsection*{Acknowledgements}

This work was supported by the Russian Foundation for Basic  Research,
project 05-01-00824.

\end{document}